\documentclass{article}
\usepackage{amsmath}                             
\usepackage{amsfonts}
\input psfig.sty
\def\fig{.}
\newtheorem{theorem}{Theorem}

\newtheorem{corollary}{Corollary}

\newtheorem{lemma}{Lemma}
\newtheorem{example}{Example}
\newtheorem{notation}{Notation}
\newtheorem{definition}{Definition}
\newtheorem{problem}{Problem}
\newtheorem{remark}{Remark}

\def\argmax{\mathop{\rm argmax}}

\def\Iden{\mbox{$\bf 1\ $}}
\def\n{\noindent}

\def\h {\mathfrak{H}}
\def\H {\mathcal{H}}
\def\p {\mathfrak{p}}

\def\g {\mathfrak{g}}
\def\k {\mathfrak{k}}

\def\so {\mathfrak{so}}
\def\su {\mathfrak{su}}

\def\Re {\mathbb{R}}

\begin{document}

\author{Navin Khaneja,\thanks{Division of Applied Sciences, Harvard University, Cambridge, MA 02138. Email:navin@hrl.harvard.edu}\ \ \ Steffen J. Glaser, \thanks{Institute
of Organic Chemistry and Biochemistry II, Technische Universit\"at M\"unchen,
85747 Garching, Germany. This work was funded by the Fonds der Chemischen Industrie and the Deutsche Forschungsgemeinschaft under grant Gl 203/1-6.}\ \ \ Roger Brockett\thanks{Division of Applied Sciences, Harvard University, Cambridge, MA 02138. This work was funded by the Army grant DAAG 55-97-1-0114, Brown Univ. Army DAAH 04-96-1-0445, and MIT Army DAAL03-92-G-0115.}}

\title{{\bf Sub-Riemannian Geometry and Time Optimal Control of Three Spin Systems: Quantum Gates and Coherence Transfer}}

\maketitle
\begin{center}
{\bf Abstract}
\end{center}
\n Radio frequency pulses are used in Nuclear Magnetic Resonance spectroscopy to produce unitary transfer of states. Pulse sequences that accomplish a desired transfer should be as short as possible in order to minimize the effects of relaxation, and to optimize the sensitivity of the experiments. Many coherence transfer experiments in NMR, involving network of coupled spins use temporary spin-decoupling to produce desired effective Hamiltonians. In this paper, we demonstrate that significant time can be saved in producing an effective Hamiltonian if spin-decoupling is avoided. We provide time optimal pulse sequences for producing an important class of effective Hamiltonians in three-spin networks. These effective Hamiltonians are useful for coherence transfer experiments in three-spin systems and implementation of indirect swap and $\Lambda_2(U)$ gates in the context of NMR quantum computing. It is shown that computing these time optimal pulses can be reduced to geometric problems that involve computing sub-Riemannian geodesics. Using these geometric ideas, explicit expressions for the minimum time required for producing these effective Hamiltonians, transfer of coherence and implementation of indirect swap gates, in a 3-spin network are derived (Theorem 1 and 2). It is demonstrated that geometric control techniques provide a systematic way of finding time optimal pulse sequences for transferring coherence and synthesizing unitary transformations in quantum networks, with considerable time savings (e.g. $42.3 \%$ for constructing indirect swap gates). 

\begin{center}
\section{Introduction}\end{center}The central theme of this paper is to compute the minimum time it takes to produce a unitary evolution in a network of coupled quantum systems, given that there are only certain specified ways we can effect the evolution. This is the problem of time optimal control of quantum systems \cite{time.opt, khaneja.thesis, herbruggen}. This problem manifests itself in numerous contexts. Spectroscopic fields, like nuclear magnetic resonance (NMR), electron magnetic resonance and  optical
spectroscopy rely on a limited set of control variables in
order to create desired unitary transformations \cite{Ernst, Science, optics}.
In NMR, unitary transformations are used to
manipulate an ensemble of nuclear spins, e.g. to transfer
coherence between coupled spins in multidimensional
NMR-experiments \cite{Ernst} or to implement quantum-logic
gates in NMR quantum computers \cite{QC}. The sequence of 
radio-frequency pulses that generate a desired unitary operator 
should be as short as possible in order to minimize the effects of 
relaxation or decoherence that are always present. In the context of quantum 
information processing, it is important to find 
 the fastest way to implement quantum gates in a given quantum technology. Given a set of universal gates, what is the most efficient way of constructing a quantum circuit given that certain gates are more expensive in terms of time it takes to implement them. All these questions are also directly related to the question of determining the minimum time required to produce a unitary evolution in a quantum system. 

\n Recall the unitary state evolution of a quantum system is given by $$ |\psi(t)> = U(t) |\psi(0)>,$$ where $|\psi(t)>$ represents the systems state vector, at some time $t$. The unitary propagator $U(t)$ evolves according to the Schr\"oedinger's equation
\begin{equation} \label{eq:unitary} \dot{U} = -i H(t)U, \end{equation} where $H(t)$ is the Hamiltonian of the system. We can decompose the total Hamiltonian as $$H = H_d + \sum_{j=1}^m u_j H_j ,$$ where $H_d$ is the internal Hamiltonian of the system and corresponds to couplings or interactions in the system. $H_j$ are the control Hamiltonians which can be externally effected \cite{khaneja}. The question we are interested in asking is, what is the minimum time it takes to drive this system \ref{eq:unitary} from $U(0)= I$ to some desired $U_F$ \cite{time.opt,khaneja.thesis}.

In \cite{time.opt,khaneja.thesis}, a general control theoretic framework for the study and design of time optimal pulse sequences in coherent spectroscopy was established. It was shown that the problems in the design of shortest pulse sequences can be reduced to questions in geometry, like computing shortest length paths on certain homogeneous spaces. In this paper, these geometric ideas are used to explicitly solve a class of problems involving control of three coupled spin $\frac{1}{2}$ nuclei. In particular, the focus is on a network of coupled heteronuclear spins. We compute bounds on the minimum time required for transferring coherence in a three spin system and derive pulse sequences that accomplish this transfer. We also derive time optimal pulse sequences producing a class of effective Hamiltonians which are required for implementation of indirect swap and $\Lambda_2(U)$ gates in context of NMR quantum computing \cite{barenco}. 

The paper is organized as follows. In the following section we recapitulate the basics of product operator formalism used in NMR. The reader familiar with the product operator formalism may skip to the next section. Section \ref{sec:main.prob} presents the main problem solved in this paper. In section \ref{sec:main.ideas}, we recapitulate the key geometric ideas required for producing time optimal pulse sequences. These ideas are developed in great detail in our work \cite{time.opt}. In section 5, we use these geometric ideas to compute the time optimal pulse sequences for producing a class of effective Hamiltonians in a network of linearly coupled heteronuclear spins. Finally these ideas are used to find pulse sequences for coherence-order selective in-phase coherence transfer in three spin system and synthesis of logic gates in NMR quantum computing. 
\begin{center}\item \section{Product Operator Basis and NMR Terminology} \label{sec:algebra} \end{center}The unitary evolution of $n$ interacting spin $\frac{1}{2}$ particles is described by an element of $SU(2^n)$, the special unitary group of dimension $2^{n}$. The Lie algebra $\su (2^n)$ is a $4^n -1$ dimensional space, identified with the space of traceless $n \times n$ skew-Hermitian matrices. The inner product between two skew-Hermitian matrix elements $A$ and $B$ is defined as $<A, B> = tr(A^{\dagger}B)$. A orthogonal basis used for this space is expressed as tensor products of Pauli spin matrices \cite{sorenson} (product operator basis). Recall the Pauli spin matrices $I_x$, $I_y$, $I_z$ defined by

\begin{eqnarray*}
\label{pauli}
I_x &=& \frac{1}{2}\left(  
\begin{array}{cc}
 0 & 1\\
1 & 0 
\end{array}
\right)
\\  
I_y &=& \frac{1}{2}\left(  
\begin{array}{cc}
 0 & -i\\
i & 0 
\end{array}
\right)\\
I_z &=& \frac{1}{2}\left(  
\begin{array}{cc}
 1 & 0\\
0 & -1 
\end{array}
\right)
\end{eqnarray*}are the generators
of the rotation in the two dimensional Hilbert space and  basis for the Lie algebra of
traceless skew-Hermitian matrices $\su (2)$. They obey the well known relations
\begin{equation}\label{eq:pauli.1}
 [I_x, \ I_y] = i I_z \; \; ; \; \; [I_y ,\ I_z] = i I_x \; \; ; \;
\;   [I_z, \ I_x] = i I_y \end{equation}
\begin{equation}\label{eq:pauli.2}
I_x^2 = I_y^2 = I_{z}^2 = \frac{1}{4}\Iden 
\end{equation}where $$ \Iden = \left(  
\begin{array}{cc}
 1 & 0\\
0 & 1 
\end{array}
\right)$$

\begin{notation} {\rm We choose an orthogonal basis $\{iB_s\}$ (product operator basis), for $\su (2^n)$ taking the form 
\begin{equation}
B_s = 2^{q-1}\prod_{k=1}^{n}(I_{k\alpha})^{a_{ks}},
\end{equation}$\alpha = x, y, or \ z$ and  
\begin{equation}
\label{tensor}
 I_{k\alpha} = \Iden \otimes \cdots \otimes I_{\alpha} \otimes \Iden,
\end{equation}
where $q$ is an integer taking values between $1$ and $n$, $I_{\alpha}$ the Pauli matrix appears in the above equation \ref{tensor} only at the $k^{th}$ 
position, and $\Iden $ the two dimensional identity matrix, appears everywhere 
except at the $k^{th}$ position. $a_{ks}$ is $1$ in $q$ of the indices and $0$ in the remaining. Note that we must have $q \geq 1$ as $q = 0$ corresponds to the identity matrix and is not a part of the algebra. }
\end{notation}

\begin{example} {\rm As an example for $n=2$ the product basis for $\su (4)$ takes the form 
\begin{eqnarray*}
q = 1 &\ & i\{I_{1x}, I_{1y}, I_{1z}, I_{2x}, I_{2y}, I_{2z} \}\\
q = 2 &\ & i\{2I_{1x}I_{2x}, 2I_{1x}I_{2y}, 2I_{1x}I_{2z} \\ 
&\ & 2I_{1y}I_{2x},2I_{1y}I_{2y}, 2I_{1y}I_{2z} \\
&\ & 2I_{1z}I_{2x},2I_{1z}I_{2y}, 2I_{1z}I_{2z}.\} 
\end{eqnarray*}}
\end{example} 

\begin{remark} {\rm It is very important to note that the expression $I_{k\alpha}$ depends on the dimension $n$. For example, the expression for $I_{2z}$ for $n=2$ and $n=3$ is 
$\Iden \otimes I_z$, and $\Iden \otimes I_z \otimes \Iden$ respectively. Also observe that these operators are only normalized for $n=2$ as

\begin{equation}\label{eq:ortho}tr(B_r B_s) = \delta_{rs}2^{n-2}
\end{equation}}
\end{remark}

\n To fix ideas, we compute one of these operators explicitly for $n=2$ 

\[I_{1z} = \frac{1}{2}\left[  
\begin{array}{cc}
1 & 0 \\
0 & -1
\end{array}
\right] \otimes \left[  
\begin{array}{cc}
1 & 0 \\
0 & 1
\end{array}
\right]   
\]  
which takes the form 
\[I_{1z} = \frac{1}{2}\left[  
\begin{array}{cccc}
1& 0 & 0 & 0\\
 0& 1 & 0 & 0\\
0& 0 & -1 & 0 \\
0 & 0 & 0 & -1
\end{array}
\right].   
\]

In this paper we want to control a network of coupled heteronuclear spins. The internal Hamiltonian for a network of weakly coupled spins takes the form
$$H_d = 2 \pi \sum_{i} \nu_iI_{iz} + 2 \pi \sum_{ij} J_{ij} I_{iz}I_{jz}.$$ Where $\nu_i$ represents Larmor frequencies for individual spins and $J_{ij}$ represents couplings between the spins. The values of the frequencies $\nu_i$ and $J_{ij}$ depend on the particular spins being used; typically, $\nu_i = 10^{8}-10^{9}$ Hz while for neighboring spins $J_{ij} = 10-10^2$ Hz. Throughout this paper, we will assume that the Larmor frequencies of spins are well separated $(|\nu_i - \nu_j| \gg |J_{ij}|)$.
 In a frame rotating about the $z$ axis with the spins at respective frequencies $\nu_i$, the Hamiltonian of the system takes the form $$H_d = 2 \pi \sum_{ij} J_{ij} I_{iz}I_{jz}.$$ 

We can also apply external radio frequency (rf) pulses on resonance to each spin. Under the assumption of wide separation of larmor frequencies, the total Hamiltonian in the rotating frame can be approximated by $$H= 2 \pi \sum_{ij} J_{ij} I_{iz}I_{jz} + 2 \pi \sum_{i}(v_{i1}I_{ix} + v_{i2}I_{iy}),$$ where $I_{ix}$ and $I_{iy}$ represent Hamiltonians that generate $x$ and $y$ rotations on the $i^{th}$ spin. By application of a resonant rf field, also called a {\it selective pulse}, we can vary $v_{i1}$ and $v_{i2}$ and thereby perform selective rotations on individual spins. In this context, we use the term {\it hard pulse} if the radio-frequency (rf) amplitude is much larger than characteristic spin-spin couplings. Such hard pulses can still be spin-selective if the frequency difference between spins is larger than the rf amplitude (measured in frequency units)\cite{Ernst}. In particular, this is always the case for the heteronuclear spins under consideration. In many situations, it is possible to ``turn off'' one or more of these couplings $J_{ij}$. This is done through standard {\it spin decoupling} techniques, for details see \cite{Ernst} and appendix A.  

We now present the main problem addressed in this paper.
\begin{center}\section{Optimal Control in Three Spin System}\label{sec:main.prob}\end{center}\begin{problem}\label{prob:main} {\rm Consider a chain of three heteronuclear spins coupled by scalar couplings ($J_{13} = 0$). Furthermore assume that it is possible to selectively excite each spin (perform one qubit operations in context of quantum computing). The goal is to produce a desired unitary transformation $U \in SU(8)$, from the specified couplings and single spin operations in shortest possible time. This structure appears often in the NMR situation. The unitary propagator $U$, describing the evolution of the system in a suitable rotating frame is well approximated by \begin{equation} \label{eq:het.three} \dot{U} = -i (\ H_d + \sum_{j=1}^{6}u_j H_j \ )U , \ \ U(0)= I \end{equation} where \begin{eqnarray*}H_d &=& 2\pi J_{12} I_{1z} I_{2z} + 2 \pi J_{23}I_{2z}I_{3z}, \\
H_1 &=& 2\pi I_{1x}, \\
\ H_2 &=& 2\pi I_{1y}, \\
\ H_3 &=& 2\pi I_{2x}, \\
H_4 &=& 2\pi I_{2y}, \\
H_5 &=& 2 \pi I_{3x}, \\
H_6 &=& 2 \pi I_{3y}.
\end{eqnarray*} The symbol $J_{12}$ and $J_{23}$ represents the strength of scalar couplings between spins $(1,2)$ and $(2,3)$ respectively. We will be most interested in a unitary propagator of the form 
$$U = \exp(-i \theta \ I_{1\alpha}I_{2 \beta}I_{3 \gamma}).$$ Where the index $\alpha, \beta, \gamma \in \{x,y,z \}$. These propagators are hard to produce as they involve trilinear terms in the effective Hamiltonian. We will refer to such propagators as {\it trilinear propogators}. To highlight geometric ideas, here we will treat the important case of this problem when the couplings are both equal ($J_{12} = J_{23} = J $). Without loss of any generality we assume $J>0$.}
\end{problem}

\begin{remark} {\rm Please note that it suffices to compute the minimum time required to produce the propagators belonging to the one parameter family $$U_F = \exp(-i \theta \ I_{1z}I_{2z}I_{3z}), \ \theta \in [0, 4 \pi],$$ because all other propagators belonging to the set $\{\exp(-i \theta \ I_{1\alpha}I_{2 \beta}I_{3 \gamma})| \alpha, \beta, \gamma \in \{x,y,z\} \}$ of trilinear propogators can be produced from $U_F$ in arbitrarily small time by selective hard pulses. As an example $$\exp(-i \theta \ I_{1x}I_{2z}I_{3z}) = \exp(-i \frac{\pi}{2}\ I_{1y})\exp(-i \theta \ I_{1z}I_{2z}I_{3z})\exp(i \frac{\pi}{2}\ I_{1y}).$$ It will be shown that finding shortest pulse sequences for these propogators, constitute an essential step in optimal implementations of logic gates in the context of NMR quantum computing.}  
\end{remark}

\begin{figure}[h]
\centerline{\psfig{file=\fig/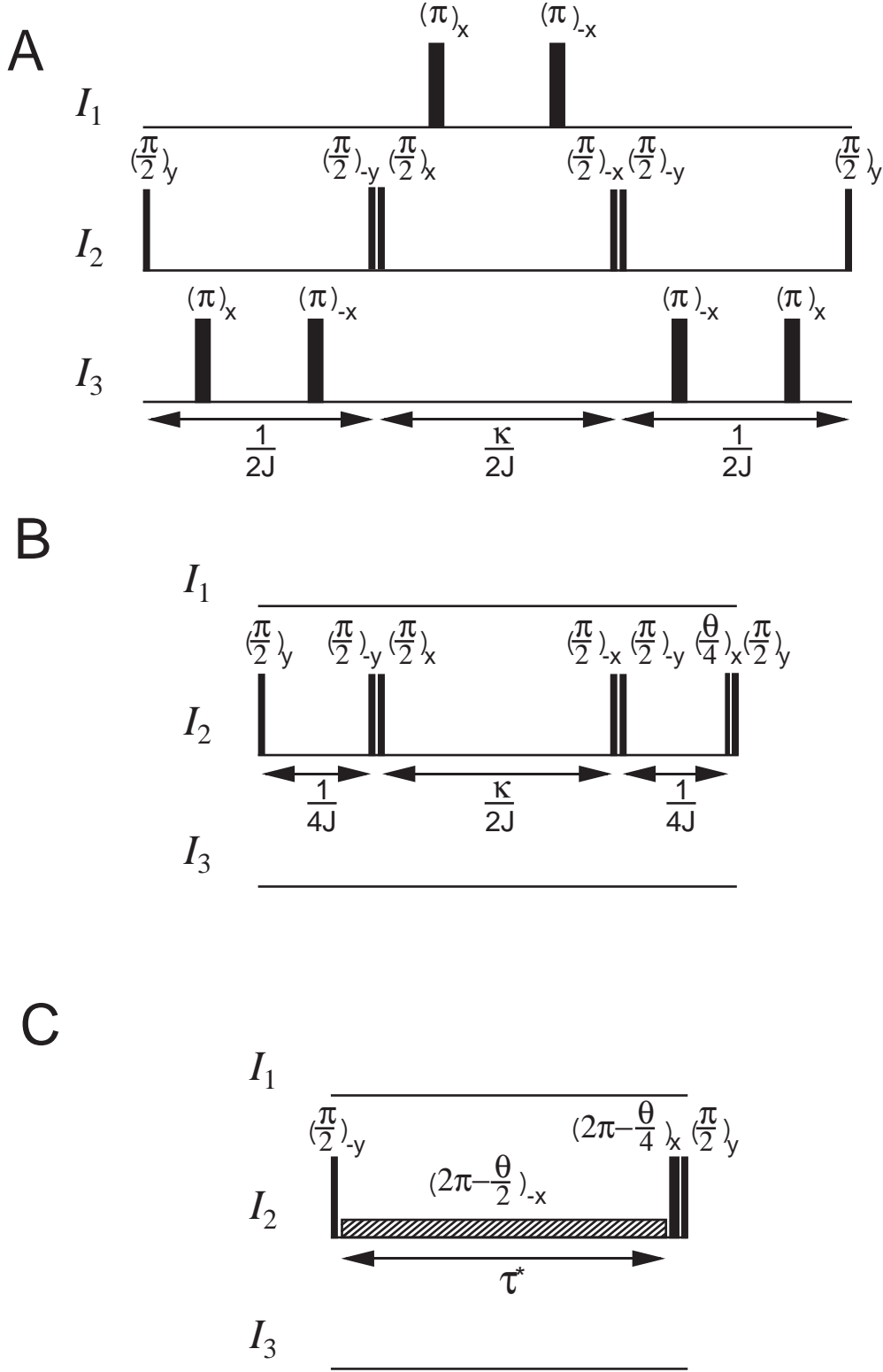}}
\caption[pulse]{The panel shows three pulse sequences for 
synthesizing the trilinear propagator 
$U_F=exp(- i \theta I_{1z} I_{2z} I_{3z})$ with
$\theta=2 \pi \kappa$. The conventional pulse sequence A uses decoupling
and takes time $t= (2+\kappa)/2J$. The second pulse sequence B improves the
first sequence by avoiding decoupling and has a duration
$t^\prime=(1+\kappa)/2J$.The final pulse sequence C is time optimal and has
a duration $t^\ast=\sqrt{\kappa(4-\kappa)}/2J$. The radio-frequency
amplitude $\nu_{rf}$ of the hatched pulse is
$(2-\kappa)J/\sqrt{\kappa(4-\kappa)}$.
\label{fig:pulse}}
\end{figure}

\n \begin{remark} \label{rem:decouple}{\rm We first compute the minimum time it takes to produce the propagator of the above type using spin-decoupling. The main computational tool used for this purpose is the Baker Campbell Hausdorff formula [BCH] \cite{Ernst}. Recall given the generators $A, B, C$ satisfying $$ [A, B] = C \ , [B, C] = A,\ [C, A] = B.$$ The BCH implies $$\exp(At)B\exp(-At) = B \cos t + C \sin t , $$ and therefore $$\exp(At)\exp(B)\exp(-At) = \exp(B \cos t + C \sin t). $$ This can be then used in problem \ref{prob:main} to produce a propagator of the form $ \exp(-i \theta \ I_{1z}I_{2z}I_{3z})$. 

\n The standard procedure uses decoupling and operates by first decoupling spin $3$ from the network (this can be achieved by standard refocusing techniques \cite{Ernst}, see Fig. \ref{fig:pulse}(A). A brief review of the basic ideas involved in spin-decoupling is presented from a control viewpoint in appendix A). The effective Hamiltonian then takes the form $$H_{eff}^1 = 2 \pi J I_{1z}I_{2z}.$$ Now by use of external rf pulses and the Hamiltonian $H_{eff}^1$, we can generate the unitary propagator $\exp(-i \pi I_{1z}I_{2x})$ as follows. 

$$ \exp(-i \frac{\pi}{2} I_{2y})\exp(-i\frac{H_{eff}^1}{2J} )\exp(i \frac{\pi}{2} I_{2y}) = \exp(-i \pi I_{1z}I_{2x}).$$

\noindent The creation of this propogator takes $\frac{1}{2J}$ units of time.

\n Similarly by decoupling spin $1$ from the network, we are left with an effective Hamiltonian $H_{eff}^2 = 2 \pi J I_{2z}I_{3z}$, which can be used along with external rf pulses to produce a propagator $\exp(-i \frac{\theta I_{2y}I_{3z}}{2})$, which takes another $\frac{\theta}{4\pi J}$ units of time. Now using the commutation relations 
$$ [2I_{1z}I_{2x}, 2I_{2y}I_{3z}] = i4I_{1z}I_{2z}I_{3z}, $$ 
$$ [4I_{1z}I_{2z}I_{3z},2I_{1z}I_{2x}] = i2I_{2y}I_{3z}, $$ 
$$ [4I_{1z}I_{2z}I_{3z}, 2I_{1z}I_{2x}] = i2I_{2y}I_{3z}.$$

\n We obtain that $$ \exp(-i \pi I_{1z}I_{2x})\exp(-i \frac{\theta I_{2y}I_{3z}}{2})\exp(i \pi I_{1z}I_{2x}) = \exp(-i \theta I_{1z}I_{2z}I_{3z}).$$ Therefore the total time required to produce the unitary propagator is $$\frac{1}{2J} + \frac{\theta}{4\pi J} + \frac{1}{2J} = \frac{4 \pi + \theta}{4 \pi J} = \frac{2 + \kappa}{2 J},$$ where $\kappa = \frac{\theta}{2 \pi}$ (see Fig \ref{fig:pulse}). }
\end{remark}

\n We will show that this propagator can be produced in a significantly shorter time using pulse sequences derived using ideas from results in geometrical control theory. 
\n Before we turn to time optimal pulse sequences, we give new implementations of the trilinear propagators that are considerably shorter than the ones given in remark \ref{rem:decouple}, even though they are not time optimal. These sequences do not involve decoupling. We present one such sequence here, for comparison with the time optimal pulse sequences in theorem \ref{th:main.0}(see Fig. \ref{fig:pulse}(B)).

\begin{notation} \label{not:main} { \rm Let $A =-i(I_{1z}I_{2x} + I_{2x}I_{3z})$, $B = -i(I_{1z}I_{2y} + I_{2y}I_{3z})$, $C = -i(2 I_{1z}I_{2z}I_{3z} + \frac{I_{2z}}{2})$ and $D = -i(4 I_{1z}I_{2z}I_{3z})$. Then observe the following commutation relations hold \begin{equation}\label{eq:so3} [A, B] = C; \ \ [B, C] = A; \ \ [C,A] = B . \end{equation} $$[A, D] = -B;\ \ [B, D ] = A. $$ }
\end{notation}

\begin{definition} \label{def:so3} {\rm Any set of three generators $A, B, C$ satisfying the equation (\ref{eq:so3}) will be referred to as the $\so(3)$ Lie algebra.} \end{definition}

\begin{figure}[t]
\centerline{\psfig{file= \fig/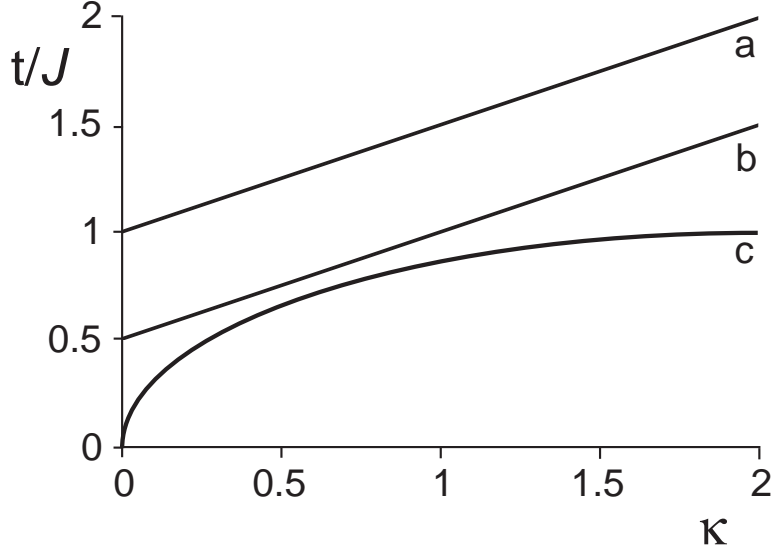 ,width=4in}}
\caption[coset]{The graph 
shows the comparison of time required by pulse sequences for creating trilinear propogators as a function of $\kappa = \theta/2\pi$. (a) Pulse sequence  using spin-decoupling, (b) improved sequence without decoupling (see remark 4), (c) time optimal pulse sequence (see theorem \ref{th:main.0}).   
\label{fig:comparison}}
\end{figure}

\begin{remark} {\rm Using the commutation relations stated above, it follows from BCH that $$P = \exp(\frac{\pi}{2}A) \exp(\frac{\theta}{2}B)\exp(- \frac{\pi}{2}A) = \exp(-i \theta (I_{1z}I_{2z}I_{3z} + \frac{I_{2z}}{4})).$$ It takes arbitrarily small time to generate the propagator $ Q = \exp(i \theta \frac{I_{2z}}{4}))$, using selective hard pulses. Thus the time required to generate the desired propagator $PQ = \exp(-i (\theta I_{1z}I_{2z}I_{3z}))$ is just the time needed to produce $P$, which can be computed explicitly. The propagator $\exp(\frac{\pi}{2}A)$ requires $\frac{1}{4J}$ units of time, and the propagator 
$\exp(\frac{\theta}{2}B)$ requires  $\frac{\theta}{4 \pi J}$ units of time. Hence the total time is $$\frac{1}{4J} + \frac{\theta}{4\pi J} + \frac{1}{4J} = \frac{1+ \kappa}{2J}.$$ Thus we see that it is possible to reduce the time of pulse sequences for implementing desired effective Hamiltonians, by not decoupling spins in the network. The savings are as much as $50\%$ for small $\kappa$ (see figure \ref{fig:comparison}) }
\end{remark}

\n We now state results on time optimal pulse sequences for coherence transfer and synthesis of logic gates in 3 spin systems. The main theorems of this paper are stated as follows. 

\begin{theorem}\label{th:main.0}{\rm \ Given the spin system in (\ref{eq:het.three}), with $J_{12} = J_{23} = J$ and $J_{13}=0$, the minimum time $t^{\ast}(U_F)$ required to produce a propagator of the form $U_F = \exp(-i \theta I_{1z}I_{2z}I_{3z}),\ \ \theta \in [0, 4 \pi] $ is given by $$t^{\ast}(U_F) = \frac{\sqrt{2\pi \theta - (\theta/2)^{2}}}{2 \pi J} = \frac{\sqrt{\kappa(4 - \kappa)}}{2J},$$ where $\kappa = \frac{\theta}{2 \pi}$.}
\end{theorem}

\n This theorem can be used to compute the minimum time and the shortest pulse sequence required for in-phase coherence transfer in the three spin network given by equation (\ref{eq:het.three}) and construction of swap gates between spin $1$ and $3$. This is stated in the following theorem.

\begin{theorem}\label{th:main.1}{\bf (Indirect Swap Gates and Coherence Transfer:)}{\rm \ Given the spin system in (\ref{eq:het.three}), with  $J_{12} = J_{23} = J$ and $J_{13}=0$, the minimum time required for producing a swap gate between spin $1$ and $3$ is $\frac{3 \sqrt 3}{2 J}$. The minimum time required for the complete in-phase transfer $I_1^{-} = (I_{1x} - iI_{1y})$ to $I_3^{-} = (I_{3x} - iI_{3y})$ is $\leq \frac{3 \sqrt 3}{2 J}$. } 
\end{theorem}

\begin{remark} {\rm The conventional approach for the above indirect swap gate involves three direct swap operations. The first operation swaps spin $1$ and $2$, followed by a swap $2$ and $3$ and finally a swap between $1$ and $2$ again. Each operation takes $\frac{3}{2J}$ units of time. The total time for this pulse sequence is $\frac{9}{2J}$. Compared to this the time optimal sequence only takes $\frac{1}{\sqrt{3}} = 57.7 \%$ of the total time. It is possible to transfer $I^{-}_1 \rightarrow I^{-}_3$ completely using two sequential selective isotropic steps that involves decoupling, each of which takes $\frac{3}{2J}$ units of time \cite{Isotropic}. This takes in total
$\frac{3}{J}$ units of time. The improved pulse sequence takes at most $\frac{\sqrt{3}}{2} = 86.6 \%$ of this time. }
\end{remark}

\n We now derive the time optimal pulse sequences that give the shortest times described in above theorems. We begin by recapitulating the main geometric ideas developed in \cite{time.opt} for finding these time optimal pulse sequences.
\begin{center}\section{Main Ideas}\label{sec:main.ideas}\end{center}
\n Let $G$ denote the unitary group under consideration. In the equation $$\dot{U} = -i(H_d + \sum_{j=1}^mv_jH_j)\ U,\ U(0)=I,$$ the set of all $U' \in G$ that can be reached from Identity $I$ within time $t$ will be denoted by ${\bf R}(I,t)$. We define\begin{eqnarray*}
t^{\ast}(U_F) &=& \inf\ \{t \geq 0 |\ U_F \in \overline{{\bf R}(I,t)} \} \\
\end{eqnarray*} where $\overline{{\bf R}(I,t)}$ is the closure of the set ${\bf R}(I,t)$, and $I$ is the identity element. $t^{\ast}(U_F)$ is called the {\it infimizing time} for producing the propagator $U_F$. Observe that the control Hamiltonians $\{H_j \}$, generate a subgroup $K$, given by $$K = \exp(\{H_j \}_{LA}),$$ where $\{H_j \}_{LA}$ is the Lie algebra generated by $\{-iH_1, -iH_2, \dots, -iH_m \}$. It is assumed that the strength of the control Hamiltonians can be made arbitrary large. This is a good approximation to the case when the strength of external Hamiltonians can be made large compared to the internal couplings represented by $H_d$. Under these assumptions the search for time optimal control laws can be reduced to finding constrained shortest length paths in the space $G/K$. It can be shown \cite{time.opt}, that 

\n \begin{theorem} \label{th:Equivalence} {\bf(Equivalence theorem):} {\rm The infimizing time $t^{\ast}(U_F)$ for steering the system $$\dot{U} = -i[H_d + \sum_{j=1}^{m}v_j H_j ] U $$ from $U(0) = I$ to $U_F$ is the same as the minimum time required for steering the adjoint system \begin{equation} \label{eq:adjoint.0} \dot{P} = \H P, \ \H \in Ad_K(-iH_d), \ P \in G \end{equation} from $P(0)=I$ to $K U_F$, where $Ad_K(-iH_d) = \{ k_1^{\dagger}(-iH_d) k_1 | k_1 \in K \}$. }
\end{theorem}

We will use this result to find time optimal pulse sequences for 3-spin system.
The key observation leading to the Equivalence theorem is summarized as follows.

\n {\bf (Minimum time to go between cosets:)} If the strength of the control Hamiltonians can be made very large, then starting from identity propagator, any unitary propagator belonging to $K$ can be produced in arbitrarily small time. This notion of arbitrarily small time is made rigorous using the concept of infimizing time as defined earlier. Therefore if $U_F \in K$ then $t^{\ast}(U_F) = 0$. Similarly, starting from $U_1$, any $kU_1,\ \ k \in K$ can be reached in arbitrarily small time. This strongly suggests that to find the time optimal controls $v_i$ which drive the evolution $(\ref{eq:unitary})$ from $U_1$ to $U_2$ in minimum possible time, we should look for the fastest way to get from the coset $KU_1$ to $KU_2$ (the coset $KU_1$ denotes the set $\{kU_1|k \in K \}$).

\n {\bf (Controlling the direction of flow in $G/K$ space:)} The problem of finding the fastest way to get between points in $G$ reduces to finding the fastest way to get between corresponding points (cosets) in $G/K$ space. Let $\g$ represent the Lie algebra of the generators of $G$ and $\k = \{H_j\}_{LA}$ represent the Lie algebra of the generators of the subgroup $K$. Consider the decomposition $\g = \p \oplus \k$ such that $\p$ is orthogonal to $\k$ and represents all possible directions in the $G/K$ space.  The flow in the group $G$, is governed by the evolution equation $(\ref{eq:unitary})$ and therefore constraints the accessible directions in the $G/K$ space. The directly accessible directions in $G/K$, are represented by the set $Ad_K(-iH_d)$. To see this, observe that the control Hamiltonians do not generate any motion in $G/K$ space as they only produce motion inside a coset. Therefore all the motion in $G/K$ space is generated by the drift Hamiltonian $H_d$. Let $k_1$ and $k_2$ belong to $K$, the coset containing identity. Under the drift Hamiltonian $H_d$, these propagators after time $\delta t$, will evolve to $\exp(-iH_d\ \delta t)k_1$ and $\exp(-iH_d \ \delta t)k_2$, respectively.  Note $$\exp(-iH_d \ \delta t)k_1 = k_1(k_1^{\dagger}\exp(-iH_d \ \delta t)k_1)$$ 
and thus is an element of the coset represented by $$ k_1^{\dagger}\exp(-iH_d  \ \delta t)k_1 = \exp (-i k_1^{\dagger}H_d k_1 \ \delta t).$$ Similarly $\exp(-iH_d \delta t)k_2$ belongs to the coset represented by element $\exp (-i k_2^{\dagger}H_d k_2 \ \delta t)$ . Thus in $G/K$, we can choose to move in directions given by $k_1^{\dagger}(-iH_d) k_1$ or $k_2^{\dagger}(-iH_d)k_2$, depending on the initial point $k_1$ or $k_2$. Therefore all directions $Ad_K(-iH_d)$ in $G/K$ can be generated by the choice of the initial $k \in K$, by use of control Hamiltonians $\{H_j \}$ (We can move in K so fast that the system hardly evolves under $H_d$ in that time). The set $Ad_K(-iH_d)$ is called the adjoint orbit of $-iH_d$ under the action of the subgroup $K$. This form of direction control has been defined as an adjoint control system \cite{time.opt}. Observe that the rate of movement in the $G/K$ space is always constant because all elements of $Ad_K(iH_d)$ have the same norm, $\|H_d\| = \|k^{\dagger}H_d k \|$ ($k$ is unitary so $kk^{\dagger}$ is identity). Therefore the problem of finding the fastest way to get between two points in the space $G/K$ reduces to finding the shortest path between those two points under the constraint that the tangent direction of the path must always belong to the set $Ad_K(-iH_d)$. This is the content of equivalence theorem.

\n {\bf (Finding Sub-Riemannian Geodesics in Homogeneous spaces:)}\ The set of accessible directions $Ad_K(-iH_d)$, in general case is not the whole of $\p$, the set of all possible directions in $G/K$. Therefore all the directions in $G/K$ space are not directly accessible. However, motion in all directions in $G/K$ space may be achieved by a back and forth motion in directions we can directly access. This is the usual idea of generating new directions of motion by using non-commuting generators (\ $\exp(\epsilon A)\ \exp(\epsilon B)\exp(-\epsilon A)\exp(-\epsilon B) \sim \exp(-\epsilon^2 [A,B])$\ ). The problems of this nature, where one is required to compute the shortest paths between points on a manifold subject to the constraint that the tangent to the path always belong to a subset of all permissible directions have been well studied under sub-Riemannian geometry. These contrained geodesics are called the sub-Riemannian geodesics \cite{brockett.singular}. The problem of finding time optimal control laws, then reduces to finding sub-Riemannian geodesics in the space $G/K$, where the set of accessible directions is the set $Ad_K(-iH_d)$. 

\n In \cite{time.opt}, these sub-Riemannian geodesics were computed for the space $\frac{SU(4)}{SU(2)\otimes SU(2)}$, in the context of optimal control of coupled 2-spin systems. It was shown that the space $\frac{SU(4)}{SU(2)\otimes SU(2)}$ has the structure of a Riemannian symmetric space which facilitates explicit computation of these constrained geodesics. In the following sections we will study these sub-Riemannian geodesics to compute the time optimal control for three spin systems. 
\begin{center} \label{sec:main.proof} \section{Time Optimal Pulse Sequences}\end{center}In the following lemma, we describe the infimizing time for the heteronuclear three spin system, described by the equation (\ref{eq:het.three}) with $J_{12} = J_{23} = J$ and $J_{13}=0$, in terms of its associated adjoint control system  $$ \dot{P} = \H P, \ \ \H \in Ad_K(-i2 \pi J (I_{1z}I_{2z} + I_{2z}I_{3z})),$$ where $K$ denotes the subgroup generated by control Hamiltonians $\{H_j \}_{j=1}^{6}$.

\begin{lemma}\label{th:op3spin.transfer}{\rm
In equation (\ref{eq:het.three}), let $K$ denote the subgroup generated by control Hamiltonians $\{H_j \}_{j=1}^{6}$. The infimizing time $t^{\ast}(U_F)$, required to produce a unitary propagator $U_F$ is the same as the minimum time $T$, required to steer the adjoint control system \begin{equation} \label{eq:adjoint} \dot{P} = \H P, \ \ \H \in Ad_K(-i 2 \pi J (I_{1z}I_{2z} + I_{2z}I_{3z})), \end{equation} from $P(0)= I$ to $P(T) \in KU_F$.} 
\end{lemma}
 
\n {\bf Proof:} The lemma follows directly from the equivalence theorem 
\ref{th:Equivalence} \hfill{\bf{Q.E.D}}.

In the following theorem, we develop a characterization of time optimal control laws for the adjoint control system (\ref{eq:adjoint.0}). This characterization is obtained using the maximum principle of Pontryagin. We briefly 
review the maximum principle here. The reader is advised to look at the reference \cite{pont} for more details.

\begin{remark} {\bf Pontryagin Maximum Principle:}  {\rm Consider the control problem of minimizing the time required to steer the control system $$\dot{x} = f(x,u),\ \ x \in \Re^{n},\ u \in \Omega \subset \Re^k, $$ from some initial state $x(0) = x_0$ to some final state $x_1$. The Pontryagin maximum principle states that if the control $\bar{u}(t)$ and the corresponding trajectory $\bar{x}(t)$ are time optimal then there exists an absolutely continuous vector $\lambda(t) \in \Re^{n}$, such that the Hamiltonian function $\h(x(t), \lambda(t), u(t)) = \lambda^{T}(t)f(x(t),u(t))$, satisfies
$$\h( \bar{x}(t), \lambda(t), \bar{u}(t)) = \max_{u \in \Omega}\h(\bar{x}(t), \lambda(t), u),$$ and $$\dot{\lambda_j}(t) = -\frac{\partial \h}{\partial x_j},\ \ j \in 1 \dots n.$$ The vector $\lambda(t)$ is called the adjoint vector and any triple $(x, \lambda, u)$ that satisfies the above conditions is called an extremal pair. The basic ideas of this theorem can be then generalized to control problems defined on Lie Groups \cite{brockett}. We use these ideas to give the necessary conditions for the time optimal control laws for the adjoint control system (\ref{eq:adjoint.0}). }   \end{remark}

\begin{theorem} \label{th:pont.max}{\rm For the adjoint control system (\ref{eq:adjoint.0}), if $\bar{\H}(t)$ is the time-optimal control law, and $\bar{P}(t)$ is the corresponding optimal trajectory, such that $\bar{P}(0)= I$ and $\bar{P}(T) \in KU_F$, then for $t \in [0, T]$, there exists $M(t) \in \p$, (directions in G/K space) such that 
\begin{eqnarray} 
\label{eq:var1} \bar \H(t) &=& \argmax_{\H} \ tr(\H M(t)),\ \ \H \in Ad_K(-iH_d), \\
\label{eq:var2} \frac{d\bar P(t)}{dt} &=& \bar \H(t) \bar P(t), \\
\label{eq:var2} \frac{d M(t)}{dt} &=& [\bar \H(t), M(t)]
\end{eqnarray} }
\end{theorem}

\n {\bf Proof:} First note $\H^{\dagger}= -\H$ as $\H$ is skew-Hermitian. We represent the linear functional on $\dot{P}$ as $\phi_{\lambda}(\dot P) = tr(\lambda^{\dagger}\H P)$ with $P\lambda^{\dagger} \in \p$ (the directions corresponding to $G/K$ space). The Hamiltonian function is then $$\h(P(t),\lambda(t),\H(t)) = tr(\lambda^{\dagger}(t)\H(t)P(t)).$$ Then the maximum principle gives  
\begin{eqnarray} 
\label{eq:U} \bar \H(t) &=& \argmax_{\H}\ tr(\H \bar{P}\lambda^{\dagger}),\ \ \H \in Ad_K(-iH_d), \\
\dot{\lambda}(t) &=& -\frac{\partial \h}{\partial P} \ = \ \bar{\H}(t)\lambda(t) 
\end{eqnarray}
Let $M(t) = \bar P(t)\lambda^{\dagger}(t)$. The differential equation for $M(t)$ is \begin{equation}\label{eq:M} \dot{M}(t) = [\bar{\H}(t),\ M(t)],\end{equation} such that $M(t) \in \p$ and the result follows.\hfill{\bf{Q.E.D}}.

\begin{remark} {\rm In the following theorem, we will use the maximum principle, to solve the time optimal problem of steering the adjoint control system (\ref{eq:adjoint}) from $P(0) = I$ to the coset $KU_F$, where $U_F = \exp(-i \theta I_{1z}I_{2z} I_{3z})$, $\theta \in [0, 4\pi]$. We hasten to add that the proof presented here only establishes that the control laws and the corresponding trajectories, given in the following theorem are extremal trajectories for the problem of time optimal control. A complete proof of optimality is beyond the scope and aim of the present paper and will be presented elsewhere. We first state a lemma which will be used in the following theorem.}
\end{remark}

\begin{lemma}\label{lem:period} {\rm Let  $A, B, C$ be as in the notation \ref{not:main}. Then 
$$\exp (2 \pi C) \exp(\alpha_1 A + \alpha_2 B + \alpha_3 C ) = I$$ for $ \sum_{i=1}^{3}\alpha_i^2 = (2 \pi)^2 $.}
\end{lemma}

\n {\bf Proof:} First note that $\exp(tA), \exp(tB), \exp(tC)$ are all periodic with period $4 \pi$ and satisfy the commutation relation $$ [A, B] = C \ , [B, C] = A,\ [C, A] = B.$$  The mapping $A \rightarrow -iI_x$, $B \rightarrow -iI_y$, $C \rightarrow -iI_z$, defines a diffeomorphism between the group $\exp \{A, B, C \}$ and $SU(2)$ given by $$\exp(\alpha_1 A + \alpha_2 B + \alpha_3 C) \rightarrow \exp(-i[\alpha_1 I_x + \alpha_2 I_y + \alpha_3 I_z]).$$ Now using the fact that if  $ \sum_{i=1}^{3}\alpha_i^2 = (2 \pi)^2$, then $\exp(-i[\alpha_1 I_x + \alpha_2 I_y + \alpha_3 I_z]) = -I$, we obtain that, if $\sum_{i=1}^{3}\alpha_i^2 = (2 \pi)^2$, then $$\exp (-i 2 \pi I_z) \exp(-i[\alpha_1 I_x + \alpha_2 I_y + \alpha_3 I_z]) = I.$$ Therefore $$\exp (2 \pi C) \exp(\alpha_1 A + \alpha_2 B + \alpha_3 C ) = I. $$ \hfill{\bf{Q.E.D}}.

\begin{theorem}\label{th:op3spin.transfer}{\rm
Let $U_F = \exp(-i \theta I_{1z}I_{2z} I_{3z})$, $\theta \in [0, 4\pi]$ and $\beta = 2 \pi - \theta/2$. The control law $$ \bar{\H}(t) = -i 2 \pi J [ (I_{1z}I_{2x} + I_{2x}I_{3z})\cos(\frac{\beta t}{T}) - (I_{1z}I_{2y} + I_{2y}I_{3z})\sin(\frac{\beta t}{T})], $$ steers the adjoint system \ref{eq:adjoint} from $P(0)= I$ to $P(T) \in KU_F$, in $$T = \frac{\sqrt{2 \pi \theta- \frac{\theta^2}{4}}}{2 \pi J} = \frac{\sqrt{\kappa (4 - \kappa)}}{2 J},$$ units of time and is time optimal.}\end{theorem}

\n {\bf Proof:} Let $A, B, C, D$ be as in the notation \ref{not:main}. Then using the commutation relations for these operators and the BCH, we can rewrite
$\bar{\H}(t)$ as $$\bar{\H}(t) = 2 \pi J \exp (-\frac{\beta C t}{T}) \ A \ \exp(\frac{\beta C t}{T}).$$ The corresponding trajectory $\bar{P}(t)$, takes the form
$$ \bar{P}(t) = \exp(-\frac{\beta C t}{T}) \exp(( \frac{\beta C}{T} + 2 \pi J A)t).$$ This can be verified by just differentiating the expression for $\bar{P}(t)$. Next observe that $\bar{P}(T) \in KU_F$. To see this note that
$$ \exp(-2 \pi C) \exp( 2 \pi J T A + \beta C) = I, $$ where $I$ is the identity matrix. This identity follows directly from the fact $(2 \pi J T)^2 + \beta^2 = (2 \pi)^2$ and lemma \ref{lem:period}. Therefore $$\bar{P}(T) = \exp(\frac{\theta C}{2}) = \exp(-i\theta (I_{1z}I_{2z}I_{3z} + \frac{I_{2z}}{4})),$$ implying $\bar P(T) \in KU_F$.   
To see that the control law $\bar{\H}(t)$ is extremal, observe for $$M(t) = -\bar{\H}(t) - \frac{\beta}{T} D,$$ the pair $(\bar{P}(t), M(t), \bar{\H}(t))$ satisfies the variation equations \ref{eq:U}, and \ref{eq:M}, of theorem \ref{th:pont.max}. To see this, recall $$\bar{\H}(t) = 2 \pi J ( A \cos(\frac{\beta t}{T}) - B \sin(\frac{\beta t}{T})) ,$$ therefore the commutation relations
$$ [A, -D] = B ,\ [B, -D] = -A,$$ imply $$[\bar{\H}, M] = \frac{2 \pi J \beta}{T}[ A \sin(\frac{\beta t}{T}) + B \cos (\frac{\beta t}{T})].$$ Furthermore $$\dot{M} =  \frac{2 \pi J \beta}{T}[ A \sin(\frac{\beta t}{T}) + B \cos (\frac{\beta t}{T})].$$ Therefore $M(t)$ satisfies the variational equation $\dot{M} = [\bar{\H}, M]$ and clearly $\bar{\H}(t)$ maximizes the function $tr(\H M(t))$ for $\H \in Ad_K(-i 2 \pi J (I_{1z}I_{2z} + I_{2z}I_{3z}))$ and $M(t) = -\bar{\H}(t) - \frac{\beta}{T} D,$. \hfill{\bf{Q.E.D}}.

\begin{corollary} {\rm Let $U_F = \exp(-i \theta I_{1\alpha}I_{2\beta}I_{3\gamma})$, $\theta \in [0, 4\pi]$ and $(\alpha, \beta, \gamma) \in (x,y,z)$. The minimum time $T$, required to steer the adjoint system from $P(0)= I$ to $P(T) \in KU_F$, is $$T = \frac{\sqrt{2 \pi \theta- \frac{\theta^2}{4}}}{2 \pi J}.$$}
\end{corollary}

\n {\bf Proof:} The proof follows from the observation that $I_{1\alpha}I_{2\beta}I_{3\gamma}$ belongs to the same coset as $I_{1z}I_{2z}I_{3z}$. Therefore the result of theorem \ref{th:op3spin.transfer} apply. 

\n {\bf Proof of Theorem \ref{th:main.0}:} The proof is now a direct consequence of the equivalence theorem \ref{th:Equivalence} and theorem \ref{th:op3spin.transfer}.

\n {\bf Geodesic Pulse Sequence:} The pulse sequence that produces the propagator $$U_F = \exp(-i \theta I_{1z}I_{2z} I_{3z}),$$ in theorem \ref{th:main.0} is as follows. $$U_F = \exp(-i \frac{\pi}{2}I_{2y}) \ \exp(-i [\pi + \frac{\beta}{2}] I_{2x}) \ \exp(T (-i 2 \pi J (I_{1z}I_{2z} + I_{2z}I_{3z}) + i \frac{\beta}{T}I_{2x})) \ \exp( i \frac {\pi}{2}I_{2y}). $$ Where $\beta$ and $T$ are as defined in the above theorem \ref{th:op3spin.transfer}. In Fig. \ref{fig:pulse}(C) a possible implementation of this geodesic pulse sequence is schematically shown. Although the simple implementation shown in Fig. \ref{fig:pulse}(C) is constrained in terms of bandwidth, it forms the basis of more broad band sequence which will be presented in a future experimental paper. 

\begin{center} \section{Indirect Swap Gates and Coherence Transfer in 3-Spin Networks} \end{center}
In this section, we will consider the problem of transfer of in-phase coherence$I_1^{-}$ to $I_3^{-}$, for the heteronuclear three spin network described by the equation (\ref{eq:het.three}). 

\begin{lemma}{ \rm The unitary propagator $$V_F = \exp(-i2 \pi (I_{1z}I_{2z}I_{3z} + I_{1y}I_{2z}I_{3y} + I_{1x}I_{2z}I_{3x})), $$ completely transfers the coherence $I_1^{-}$ to $I_3^{-}$. }
\end{lemma}

\n {\bf Proof:} First observe that $I_{1z}I_{2z}I_{3z}$, $I_{1y}I_{2z}I_{3y}$, and $I_{1x}I_{2z}I_{3x}$ commute, therefore 
$$V_F = \exp(-i2 \pi I_{1z}I_{2z}I_{3z})\  \exp(-i2 \pi I_{1y}I_{2z}I_{3y}) \ \exp(-i2 \pi I_{1x}I_{2z}I_{3x}).$$ Furthermore, observe that $\{ I_{1x},\ 4I_{1y}I_{2z}I_{3z},\ 4I_{1z}I_{2z}I_{3z} \}$ forms a $\so(3)$ Lie algebra. Therefore, $$ \exp(-i\frac{\pi}{2} (4I_{1z}I_{2z}I_{3z})) \ I_{1x} \ \exp(i\frac{\pi}{2} (4I_{1z}I_{2z}I_{3z})) = 4I_{1y}I_{2z}I_{3z}. $$ 
Also note that $\{ 4I_{1y}I_{2z}I_{3z},\ 4I_{1y}I_{2z}I_{3y}, I_{3x} \}$ forms a $\so(3)$ Lie algebra. Therefore $$ \exp(-i\frac{\pi}{2} (4I_{1y}I_{2z}I_{3y}))  \ 4I_{1y}I_{2z}I_{3z} \ \exp(i\frac{\pi}{2} (4I_{1y}I_{2z}I_{3y})) = I_{3x}. $$ Combining the above equalities we obtain $V_F I_{1x} V_F^{\dagger} = I_{3x}$. Similarly one can verify that $V_F I_{1y} V_F^{\dagger} = I_{3y}$. Hence the proof. \hfill{\bf{Q.E.D}}. 

\n {\bf Proof of Theorem \ref{th:main.1}:} {\bf (Coherence Transfer:)} We need to compute the minimum time required to produce the propagator
$$V_F = \exp(-i2 \pi I_{1z}I_{2z}I_{3z})\ \exp(-i2 \pi I_{1y}I_{2z}I_{3y}) \ \exp(-i2 \pi I_{1x}I_{2z}I_{3x}).$$ We have already shown that the minimum time required to produce a propagator of 
the form $\exp(-i 2 \pi I_{1 \alpha}I_{2 \beta}I_{3 \gamma})$, where 
$(\alpha, \beta, \gamma) \in (x, y, z)$ is $$\frac{\sqrt{2\pi (2 \pi) - (\pi)^{2}}}{2 \pi J} = \frac{\sqrt{3}}{2 J}.$$  Therefore $V_F$ can be produced in time less than or equal to $\frac{3 \sqrt{3}}{2 J}$ (see following remark). Since there might be other unitary propogators, that might achieve this coherence transfer and take less time to synthesize, we can only claim that the minimum time required to transfer the coherence $I^{-}_1$ to $I^{-}_3$ is less than or equal to $\frac{3 \sqrt{3}}{2J}$.  

\n {\bf Pulse Sequence:} The pulse sequence that produces the propagator $$V_F = \exp(-i2 \pi (I_{1z}I_{2z}I_{3z} + I_{1y}I_{2z}I_{3y} + I_{1x}I_{2z}I_{3x})),$$ is as follows. Let $U_1 = \exp(-i2 \pi (I_{1z}I_{2z}I_{3z}))$, $U_2 = \exp(-i2 \pi (I_{1y}I_{2z}I_{3y}))$ and $U_3 = \exp(-i2 \pi (I_{1x}I_{2z}I_{3x}))$. Then $$ U_1 = \exp(-i \frac{\pi}{2}I_{2y}) \ \exp(-i [\pi + \frac{\beta}{2}]I_{2x}) \ \exp(T (-i 2 \pi J (I_{1z}I_{2z} + I_{2z}I_{3z}) + i \frac{\beta}{T}I_{2x})) \ \exp( i \frac{\pi}{2}I_{2y}). $$ 
$$ U_2 = \exp(i \frac{\pi}{2}I_{1x}) \ \exp(i \frac{\pi}{2}I_{3x}) \ U_1 \ \exp(-i \frac{\pi}{2}I_{1x}) \ \exp(-i \frac{\pi}{2}I_{3x}). $$
$$ U_3 = \exp(-i \frac{\pi}{2}I_{3y}) \ \exp(-i \frac{\pi}{2}I_{1y}) \ U_1 \ \exp(i \frac{\pi}{2}I_{1y}) \ \exp(i \frac{\pi}{2}I_{3y}). $$
Finally  $$V_F = U_1\ U_2 \ U_3. $$ Where $\beta = -\pi$ and $T = \sqrt{3}/2J$.

\begin{remark}{\rm It can in fact be shown, that the minimum time required to produce the propogator $V_F$ in the above theorem is 
$\frac{3 \sqrt{3}}{2 J}$. A rigorous proof is beyond the goals of 
the present paper, however the key observation is that, $I_{1z}I_{2z}I_{3z}$, 
$I_{1y}I_{2z}I_{3y}$, and $I_{1x}I_{2z}I_{3x}$ commute, therefore the minimum time required to produce the propagator $$V_F = \exp(-i2 \pi I_{1z}I_{2z}I_{3z})\ \exp(-i2 \pi I_{1y}I_{2z}I_{3y}) \ \exp(-i2 \pi I_{1x}I_{2z}I_{3x}), $$ is the sum of minimum time required to produce the individual 
propagators $\exp(-i2 \pi I_{1z}I_{2z}I_{3z})$, $\exp(-i2 \pi I_{1y}I_{2z}I_{3y})$ and $\exp(-i2 \pi I_{1x}I_{2z}I_{3x})$.}
\end{remark}  
 
\n {\bf Proof of Theorem \ref{th:main.1}:}\ {\bf(Indirect Swap Gates)} The indirect swap gate $U_{sw}(1,3)$ is given by
$$ U_{sw}(1,3) = \exp(-i2 \pi (I_{1z}I_{2z}I_{3z} + I_{1y}I_{2z}I_{3y} + I_{1x}I_{2z}I_{3x})) \exp(i \frac{\pi}{2}I_{2z}). $$ The propagator $\exp(i \frac{\pi}{2}I_{2z})$ can be produced in arbitrarily small time by selective hard pulses. Therefore the minimum time required to produce the swap gate is the same as the minimum time required for 
creating $\exp(-i2 \pi (I_{1z}I_{2z}I_{3z} + I_{1y}I_{2z}I_{3y} + I_{1x}I_{2z}I_{3x}))$, which is $\frac{3 \sqrt{3}}{2 J}$. Hence the proof \ \hfill{\bf{Q.E.D}}. 

\begin{remark} {\bf Synthesis of $\Lambda_2(U)$ gates:} {\rm Pulse sequences for produce $\Lambda_2$ gates, in the context of NMR quantum computing need to synthesize effective Hamiltonians of the form $I_{1\alpha}I_{2\beta}I_{3 \gamma}$. To see this, observe that $$\Lambda_2(I_z) = \left[  
\begin{array}{cccccccc}
1& 0 & 0 & 0 & 0 & 0 & 0 & 0 \\
 0& 1 & 0 & 0 & 0 & 0 & 0 & 0 \\
0& 0 & 1 & 0 & 0 & 0 & 0 & 0 \\
0 & 0 & 0 & 1 & 0 & 0 & 0 & 0 \\
0 & 0 & 0 & 0 & 1 & 0 & 0 & 0 \\
 0& 0 & 0 & 0 & 0 & 1 & 0 & 0 \\
0& 0 & 0 & 0 & 0 & 0 & 1 & 0 \\
0 & 0 & 0 & 0 & 0 & 0 & 0 & -1 \\
\end{array}
\right]. $$ This can be rewritten as $$\Lambda_2(I_z) = \exp (-i \pi \left[  
\begin{array}{cc}
0 & 0 \\
0 & 1
\end{array}
\right]\otimes \left[\begin{array}{cc}
0 & 0 \\
0 & 1
\end{array}
\right]\otimes \left[  
\begin{array}{cc}
0 & 0 \\
0 & 1
\end{array}
\right]) = \exp[-i \pi(\frac{\Iden}{2} - I_{1z})\otimes(\frac{\Iden}{2} - I_{2z})\otimes (\frac{\Iden}{2} - I_{3z})].$$ Thus the effective Hamiltonian takes the form \begin{eqnarray*} 
H_{eff} &=& \pi(\frac{\Iden}{2} - I_{1z})\otimes(\frac{\Iden}{2} - I_{2z})\otimes (\frac{\Iden}{2} - I_{3z}) \\ 
\ &=& \pi(\frac{\Iden}{8} + \frac{(I_{1z} + I_{2z} + I_{3z})}{4} +  \frac{(I_{1z}I_{2z} + I_{2z}I_{3z} + I_{1z}I_{3z})}{2} + I_{1z}I_{2z}I_{3z}).\end{eqnarray*} Since the term $I_{1z}I_{2z}I_{3z}$ commutes with other terms in the effective Hamiltonian, it needs to be produced besides the other terms in the $H_{eff}$ to synthesize the $\Lambda_2(I_z)$ gate. We have already computed the time optimal pulse sequences for the optimal implementation of an effective Hamiltonian of the form $I_{1z}I_{2z}I_{3z}$. Therefore to derive optimal implementations of $\Lambda_2(I_z)$ gates, further work is required to compute is shortest pulse sequences for synthesizing an effective Hamiltonian of the form $I_{1z}I_{3z}$.}
\end{remark}
\begin{center} \section{Conclusion} \end{center}
In this paper we have demonstrated substantial improvement in the time that is required to synthesize an important class of unitary transformations in spin systems consisting of three spins $\frac{1}{2}$. It was shown that computing the time-optimal way to transfer coherence in a coupled spin network can be 
reduced to problems of computing sub-Riemannian geodesics \cite{brockett.singular}. These problems were then explicitly solved for a linear three spin 
chain. These ideas are not just restricted to the 3-spin case considered in this paper but can be extended to find time optimal pulse sequences in a general quantum network \cite{quant-network}. 

\begin{center}
\begin{table}{\bfseries Table 1: Comparison of Pulse Sequence Durations}
\begin{tabular}{cccc}
\hline 
Unitary Transformation & $\tau$(State of the art sequences) & $\tau^{\ast}$ (Geodesic sequences) & $\frac{\tau^{\ast}}{\tau}$ \\
\hline \\
$U_F = \exp(-i 2 \pi \kappa I_{1\alpha}I_{2\beta}I_{3\gamma})$ &$\frac{2 + \kappa}{2 J}$ & $\frac{\sqrt{|\kappa|(4 - |\kappa|)}}{2|J|}$ & $\frac{\sqrt{|\kappa|(4 - |\kappa|)}}{2 + \kappa}$ \\
$U_F = \exp(-i 2 \pi I_{1\alpha}I_{2\beta}I_{3\gamma})$ &$\frac{3}{2 J}$ & $\frac{\sqrt 3}{2|J|}$ & $\frac{1}{\sqrt 3} = 57.7 \%$ \\ 
Swap(1,3) & $\frac{9}{2J}$ & $\frac{3 \sqrt 3}{2J}$ & $\frac{1}{\sqrt 3} = 57.7 \%$ \\
$I_1^{-} \rightarrow I_3^{-}$ & $\frac{3}{J}$ & $\frac{3 \sqrt 3}{2J}$ & $\frac{\sqrt 3}{2} = 86.6 \% $ \\
\hline
\end{tabular}
\end{table} 
\end{center}

%\begin{center}\label{app:decoupling}\section{Appendix:Spin-decoupling}\end{cen%ter}
\appendix
\label{app:decoupling}
\begin{center}\section{Appendix: Spin-decoupling}\end{center}
Given the evolution of the unitary propogator $$\dot{U} = -i(H_d + \sum_{j=1}^mv_jH_j)\ U,\ U(0)=I,$$ let $H_d$ have a decomposition $H_d = H_d^A + H_d^B$ such that $[H_d^A , H_d^B] = 0$. The control Hamiltonians $\{H_j\}$, generate a subgroup $K$, given by $$K = \exp(\{H_j \}_{LA}),$$ where $\{H_j \}_{LA}$ is the Lie algebra generated by $\{-iH_1, -iH_2, \dots, -iH_m \}$. Let $k \in K$ be such that 
\begin{equation}
\label{eq:decoupling}
k^{-1}-i(H_d^A + H_d^B)k = -i(H_d^A - H_d^B).
\end{equation} It is assumed that the strength of the control Hamiltonians can be made arbitrary large. Under this assumption the propogator $k$ can be produced in arbitrarily small time, 
such that the evolution due to the drift $H_d$ 
during this time can be neglected. Now consider the evolution  
$$ U(t) = \exp (-iH_d \frac{t}{2})\ k^{-1}\ \exp(-iH_d\frac{t}{2})\ k.$$ From equation \ref{eq:decoupling}, we obtain 
$$U(t) = \exp (-i[H_d^A + H_d^B] \frac{t}{2})\ \exp(-i[H_d^A - H_d^B] \frac{t}{2}) = \exp (-iH_d^A t).$$ Therefore the net evolution is 
as if the system evolved under the drift term $H_d^A$ for time 
$t$. We will say that the $H_d^B$ part of the drift has been decoupled. In 
a network of coupled spins, $H_d^B$ represents the coupling of 
a specified spin to the rest of the network and decoupling $H_d^B$ 
corresponds to decoupling the spin from the network.

\end{document}